\documentclass[a4paper,UKenglish,cleveref, autoref, thm-restate]{lipics-v2021}

\title{Improving Pedestrians Traffic Priority via Grouping and Virtual Lanes in Shared Spaces} 

\titlerunning{Improving Pedestrian Priority via Grouping and Virtual Lanes} 

\author{Yao Li\footnote{corresponding author}}{Institute of Cartography and Geoinformatics, Leibniz Universität Hannover, Germany \and \url{https://www.ikg.uni-hannover.de/de/institut/personenverzeichnis/li/} }{yao.li@ikg.uni-hannover.de}{https://orcid.org/0000-0002-5716-6568}{Research supported by the Deutsche Forschungsgemeinschaft (DFG; German Research Foundation) - 227198829/GRK1931}

\author{Vinu Kamalasanan}{Institute of Cartography and Geoinformatics, Leibniz Universität Hannover, Germany  \and \url{https://www.ikg.uni-hannover.de/en/institute/search-for-persons/kamalasanan/}}{vinu.kamalasanan@ikg.uni-hannover.de}{}{Research supported by DAAD Graduate School Scholarship Programme (GSSP)}

\author{Mariana Batista}{Institute of Transportation and Urban Engineering, Technische Universität Braunschweig, Germany  \and \url{https://www.tu-braunschweig.de/en/ivs/institut/team/batista}}{m.batista@tu-braunschweig.de}{}{Research supported by DAAD Graduate School Scholarship Programme (GSSP)}

\author{Monika Sester}{Institute of Cartography and Geoinformatics, Leibniz Universität Hannover, Germany \and \url{https://www.ikg.uni-hannover.de/en/institute/search-for-persons/sester/}}{monika.sester@ikg.uni-hannover.de}{https://orcid.org/
0000-0002-6656-8809}{}

\authorrunning{Y. Li, V. Kamalasanan, M. Batista and M. Sester} 

\Copyright{Yao Li, Vinu Kamalasanan, Mariana Batista, and Monika Sester} 



\category{Short Paper} 

\relatedversion{} 

\funding{}

\nolinenumbers 

\EventEditors{Toru Ishikawa, Sara Irina Fabrikant, and Stephan Winter}
\EventNoEds{3}
\EventLongTitle{15th International Conference on Spatial Information Theory (COSIT 2022)}
\EventShortTitle{COSIT 2022}
\EventAcronym{COSIT}
\EventYear{2022}
\EventDate{September 5--9, 2022}
\EventLocation{Kobe, Japan}
\EventLogo{}
\SeriesVolume{240}
\ArticleNo{27}

\begin{document}

\maketitle

\begin{abstract}
The shared space design is applied in urban streets to support barrier-free movement and integrate traffic participants (such as pedestrians, cyclists and vehicles) into a common road space. Regardless of the low-speed environment, sharing space with motor vehicles can make vulnerable road users feel uneasy. Yet, walking in groups increases their confidence as well as influence the yielding behavior of drivers. Therefore, we propose an innovative approach to support the crossing of pedestrians via grouping and project the virtual lanes in shared spaces. This paper presents the important components of the crowd steering system, discusses the enablers and gaps in the current approach, and illustrates the proposed idea with concept diagrams.
\end{abstract}

\section{Introduction}
\label{sec:introduction}

Shared spaces are mixed traffic environments that aim to minimize traffic control and rely on negotiation-based movement to integrate road users into a common road space \cite{Karndacharuk2014}. In the absence of traffic control measures, pedestrians are obliged to assess the situation, especially by establishing eye contact with others, before deciding to cross \cite{su11236713, Karndacharuk2014}. Other implicit and explicit communication, hand gestures being an example of it, are also used to negotiate the right of way among road users. However, even when such design settings and negotiations are expected to increase safety \cite{Karndacharuk2014}, on-street interviews and surveys indicate that vulnerable road users are not necessarily confident in sharing space with motor vehicles \cite{MoodyMelia, Hammond}.
 
From an urban planning perspective, designing shared spaces focuses on enhancing pedestrian movement and their perception of domain compared to conventional layouts. The placement of street furniture and design elements, such as a continuous shared level surface, increase the user perception of pedestrian domain and their consequent sense of priority over cars \cite{ruiz2017shared}. Although \cite{Karndacharuk2014} argued that pedestrians feel more assertive and navigate with more control in shared spaces, this unusual setting can be stressful for road users who do not feel confident sharing space with motor vehicles, leading to confusion regarding priority and directly influencing their crossing behavior. It is then crucial to address traffic conflicts arising from these particular situations. 

As mentioned in \cite{su11236713}, crossing in groups can positively impact crossing assertiveness and vehicle yielding behavior. Therefore, we propose an innovative approach to help support pedestrian movement using virtual lanes and group formation aiming to improve the confidence and safety of vulnerable road users when crossing a shared space.

\section{Background}
\label{sec:background}
\subsection{Crowd steering in public spaces}
Crowd steering focuses on the problem of steering the movement of people in public and urban spaces \cite{sassi2015}, i.e. suggesting people where to move to eventually achieve a desirable global configuration with regard to crowd distribution. It plays an important role in addressing a few real-life problems, e.g. indoor/outdoor evacuation, avoiding peculiar situations (moving across unsafe neighborhoods or extremely polluted streets), visiting a range of attractions with an optimized order, etc. 

Most crowd steering applications can be divided into the following steps: firstly, the basic dynamic data like position and orientation are collected by sensors; based on the observed data, applications are able to track multiple persons, predict future trajectories, plan paths or identify groups to accommodate various requests; the desired results are then presented to steered agents via suitable infrastructures. In the following, we listed some recently emerged technologies that enable crowd steering:

\textbf{Multiple object tracking and trajectory prediction} the task of Multiple Object Tracking (MOT) is largely partitioned into locating multiple objects, maintaining their identities, and yielding their individual trajectories given an input video \cite{luo2021multiple}. Accurate trajectory prediction is a crucial task in different communities, it enables an intelligent system to forecast the behavior of road users based on the behavior to date and make a reasonable and safe decision for the next operation \cite{hao2021}.

\textbf{Grouping and group identification} groups are often found during pedestrian movements. Here, group is not only restricted to social groups (friends or families) but also contains regulation-based groups (e.g. pedestrians following the same phase of traffic lights). Grouping increases the safety during movements by holding a buddy system \cite{su11236713}. There has been an increasing amount of literature on grouping in recent years: \cite{huang2014path} controls the merging and splitting of a single-line group by penalty; \cite{hao2019} used a time-sequence DBSCAN that based on coexisting time and Euclidean distance between pedestrians to detect groups; \cite{yao2021} groups the road user based on similar origin and destination (OD) when they enter a shared space.

Group identification in an intelligent crowd steering environment could potentially help add individuals to existing groups while new or larger groups are formed in the process. Identifying people connected to each other through forms of common interactions (e.g engaging in similar activities or goals) could help identify potential groups. This would require anticipating every individual's action and intention way ahead of time. Recently large-scale spatio-temporal individual action and social group annotation datasets \cite{ehsanpour2021jrdb} provide precise data for such identification. If the spatio-temporal information of the tracked pedestrian groups is available, group surfing approaches \cite{du2019group} could be used to increase group size over different time windows. In \cite{du2019group}, a robot-based navigation approach with sub-goals was used to promote groups, which could easily be extended to pedestrians.

\textbf{Infrastructures for displaying result} recently, researchers have shown an increasing number of physical attempts to lead pedestrians to desirable global configurations, such as dynamic signages\cite{Langner2014} and gaze-based approaches \cite{giannopoulos2015gazenav}.

\subsection{Visual augmentation to influence pedestrian walking behavior}

Enabling potential grouping behavior could be achieved using visual augmentation. Such systems could either be via projecting virtual lanes or using Augmented reality (AR) wearable devices. While projection-based approaches would require the installation of infrastructure to enable this, AR glasses like the Hololens could also be used to visualize virtual content. 

An AR-based interface presenting virtual lanes was prototyped \cite{hesenius2018don} for pedestrians which displayed the path and traffic directly into the visual field. A dynamic projection system was used in \cite{busch2018generation} where pedestrians were detected using the LIDAR sensor mounted on the vehicle, which was then used to show virtual crossing lanes. To prototype, a virtual traffic infrastructure for shared spaces \cite{kamalasanan2020behavior} demonstrated how virtual signals can control behavior using wearable AR.


Considering all of the above, we propose a concept that can be applied in crowded situations, or an individual alone situation would have difficulties getting the right of way. In the Chapter \ref{sec:system_concept_idea}, we explain our idea in a context of shared space.

\section{Gaining priority via collaboration using virtual lanes}
\label{sec:system_concept_idea}

\label{sec:system-workflow}
\begin{figure}[h]
    \centering
    \includegraphics[width=1\textwidth]{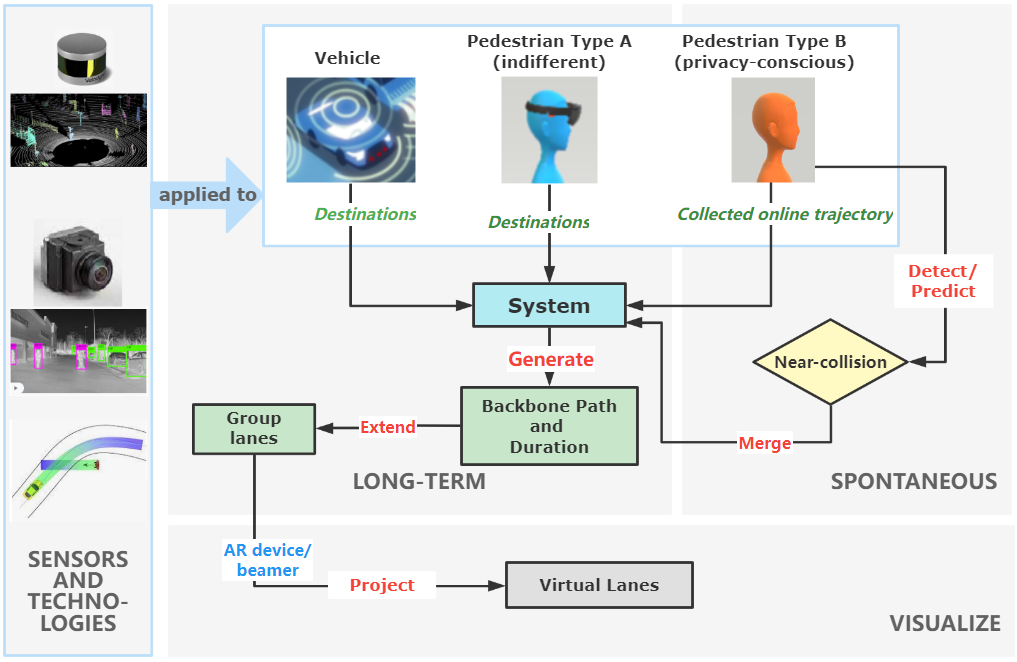}
    \caption{Concept of shared space pedestrian grouping supported by virtual lanes}
    \label{fig:system_workflow}
\end{figure}

Our concept idea of enabling pedestrian priority via augmentation and grouping is more focused on an intelligent traffic environment \cite{hu2020review}, where the traffic scene is controlled in real-time with sensors and projector displays. While sensors like camera and LIDAR enable live traffic monitoring, projection systems will dynamically show the traffic control signals, such as zebra crossing and virtual pedestrian lanes on the ground to guide/support pedestrians to cross and vehicles to give way once conflicts might happen.

As different traffic agents, for instance, autonomous taxis (ride-sharing or ride-hailing), are potentially interconnected in such an environment, they would share their OD data to a central system which could further enhance traffic signaling and real-time collision management. Our concept focuses on including pedestrian safety and priority to real-time traffic management by incorporating collaborative pedestrian motions. Collaborative pedestrian motions in this context mean pedestrians that are willing to walk together as they are moving to a similar destination \cite{bhowmick2021exploring}. This could also include shared walking  approaches if a connected systems are in place.
We also made a strong assumption that few/all pedestrians would wear AR glasses for safety and be included in the connected traffic management ecosystems, because current AR devices already dispose of the functionalities of capturing environmental information and projecting recommended routes.
Since sensors can collect dynamic data from road users and then send it to the central system, tracking and prediction algorithms could be applied in the collected data. Meanwhile, the subset of pedestrians who wear connected AR glasses could also share their walking paths and potential crossing points with the central system. We further distinguish the road users according to their privacy settings:

\textbf{Spontaneous collaborative approach} particularly considers the privacy-conscious road users (Type B) who do not share their destination information with the ecosystem. Once pedestrians are detected by installed sensors, the future motions are predicted to identify overlapping target pedestrian destinations. If the traffic system identifies multiple people walking towards the same goal, people would be motivated to walk/cross together by projecting a virtual lane/zebra crossing based on the predicted context. However, if it identifies a conflict, projected control to avoid collisions would enable centralized management of conflicts. 

\textbf{Long-term collaborative approach} is suitable for the road users who are willing to share information with the system (so-called "indifferent road users", Type A). These kinds of pedestrians is considered wearing AR glasses or other means also disclose their walking path. Hence the destination is first shared with the central system. As the system is aware of the shared OD data, it collects the road users who have similar ODs at a specific duration (waiting time), forms groups, and gives the crossing priority to larger groups to avoid conflicts and improve efficiency. Group formation can be calculated using an adapted facility location algorithm \cite{yao2021}, or edge bundling \cite{auber2010tulip}. 

To realize virtual walking/crossing paths, the system will calculate the location and duration of each of the routes and will create a path geometry. This process will take the width of a single road user unit (called backbone path) and group size into account \cite{Kamphuis2004}. Figure \ref{fig:system_workflow} provides an overview of what we envision for this ecosystem.

With our concept idea, we intend to empower pedestrians in shared traffic spaces with 1) groups creating (visual) impact and thus enforcing priority in conflicting scenarios and 2) realizing shared walking and grouping to make it more attractive for other road users to join in and thus take advantage of the joint forces.

\section{Discussion}

Some of the potential prerequisites and challenges for the realization of this approach are discussed below.

\textbf{Important parameters for grouping} pedestrians would have to gather at several suitable spots (\textit{group center}) to jointly cross the shared space. They might wait till a sufficiently large group (\textit{group size}) is formed, or till a critical \textit{waiting time} is reached. Waiting time is critical because it protects small groups from unreasonable delays, e.g., isolated pedestrian who do not belong to any groups is considered as a single-member group. Since the system tries to give larger groups priority in crossing, single-member-group may need more time to cross if there is no waiting time limit. The group center would be identified based on the \textit{distance difference} between OD of group members.

\textbf{Acceptance of virtual infrastructure for crossing} 
from a traffic planning perspective, setting gap acceptance, crossing speed and duration for virtual lanes to match motor vehicles' speed and yielding behavior can objectively indicate the potential direction and duration of the proposed visualization. Therefore, the acceptance would likely increase with virtual infrastructure that ensures a sense of safety. For that, it should integrate a critical gap acceptance that would project the virtual lane in a suitable distance in space and time from motor vehicles. 


\textbf{Impediment of remote controlling pedestrians} \cite{ZHOU2009491} claims that people had stronger intentions to illegally cross as groups when they are not given the right of the way (e.g. wait for over 90s). Thus, similar to the obedience to existing traffic signs, road users would follow the recommended virtual lanes once they experience the benefits of gaining the right of the way. Another concern comes from losing freedom while crossing, as pedestrian locomotion provides more degrees of freedom in terms of how individuals can move. Our approach is aiming to offer a safer and more efficient choice in busy traffic contexts and not limit them. 
 
\textbf{Impediments of privacy} low-resolution sensors are sufficient for tracking and prediction algorithms (e.g. in \cite{robicquet2016learning}), so individuals are not necessarily identified while crossing. Moreover, the collected data would be dropped once the crossing is completed.

\textbf{Virtual Infrastructures} early studies on applying projection-based augmentation by inducing vection \cite{furukawa2011vection} have been influential in controlling the behavior of pedestrians. But when such approaches are applied to a larger scale, for example in an outdoor shared space, their effectiveness would depend on the maturity of both wearable AR and projection systems. Such virtual lanes should be visually appealing both during the day and night to serve as effective pedestrian infrastructure. Furthermore, we also believe that AR glasses might become common in near future as smartphones today and perhaps become a mobility aid \cite{vinu_yu}, which can further support the maturity of our concept towards safety.

\textbf{Trust} the acceptance of our idea depends on the trust of pedestrians in the proposed system, especially vulnerable road users, such as elderly pedestrians who are likely to be less tech-savvy, thus refuse to wear advanced devices. However, pedestrians tend to gather while crossing, thus those pedestrians can also be attracted by the formed groups to follow the virtual lanes. According to \cite{yamaguchi2011you}, pedestrians belonging to the same group tend to automatically adjust the walking speed, making it easier for the vulnerable road users to continue crossing with the group.

\textbf{Robustness} misprediction and mis-grouping could happen. However, the user is self-sufficient to judge and make decisions in these situations. Ultimately, the proposed approach is to support vulnerable road users and not bind them. If it leads in the wrong direction, users can simply ignore it. If they still accept it, they will eventually prolong their crossing movement in the shared space. As an alternative to physical infrastructure in shared spaces, pedestrians are informally getting priority from motor vehicles when using virtual lanes, which complement social protocols such as eye contact or a brief gesture.

\textbf{User study and assessment} a user study to assess the proposed idea's acceptability and investigate objective and subjective parameters in terms of efficiency, safety, and comfort are fundamental in this case. It is paramount to understand users' perceptions of this approach and how they feel crossing such shared environments in a group and through virtual infrastructure to evaluate acceptability. Moreover, analyzing critical waiting time and gap acceptance can further improve the virtual infrastructure design to assure compliance.

\section{Conclusion and Outlook}
Our approach can help to dynamically and temporally sort and separate the mixed traffic into several virtual lanes, in order to allow for a swift movement of all participants. We also discussed the current research gaps in accomplishing such an approach. 

We believe that the challenges around our approach can be addressed more broadly, especially within the spatial information theory research, which can be of interest to the COSIT community for the related field. In future work, we plan to develop a concrete methodology to define and apply pedestrian group formation in shared spaces and evaluate the acceptability of virtual infrastructure. With that, we expect to get insights into and explore solutions for vulnerable road users in urban mixed traffic environments.



\bibliography{lipics-v2021-sample-article}

\end{document}